\documentstyle[prl,twocolumn,aps,epsf]{revtex}

\begin{document}

\title{Josephson-Junction Qubits with Controlled Couplings}

\author {Yuriy Makhlin$^{*,\dag}$, Gerd Sch\"on$^{*}$, and Alexander 
Shnirman$^{\ddag}$}

\address{
$^*$Institut f\"ur Theoretische Festk\"orperphysik,
Universit\"at Karlsruhe, D-76128 Karlsruhe, Germany. \\
$^\dag$Landau Institute for Theoretical Physics,
Kosygin St. 2, 117940 Moscow, Russia.\\
$^\ddag$Department of Physics, University of Illinois at Urbana-Champaign,
Urbana, IL 61801-3080, U.S.A.
}

\maketitle
\begin{abstract}
{\bf
Low-capacitance Josephson junctions, where Cooper pairs tunnel
coherently while Coulomb blockade effects allow the control of the
total charge, provide physical realizations of quantum bits (qubits),
with logical states differing by one Cooper-pair charge on an island.
The single- and two-bit operations required for quantum computation
can be performed by applying a sequence
of gate voltages.  A basic design, described earlier~\cite{SSH},
 is sufficient to demonstrate
the principles, but requires a high precision time control, and
residual two-bit interactions introduce errors. Here we
suggest a new nano-electronic design, close to ideal, where the
Josephson junctions 
are replaced by controllable SQUIDs. This relaxes the requirements on
the time control and  system parameters substantially,
and the two-bit coupling can be switched exactly between
zero and a non-zero value for arbitrary pairs. 
The phase coherence time is sufficiently long to allow a series of 
operations.
}\\

\end{abstract}


A quantum computer can perform certain tasks
which no classical computer is able to do in acceptable 
times~\cite{LloydIntro,Bennett,DV,Barenco}. 
It is composed of a (large) number of coupled two-state 
quantum systems forming
qubits;  the computation is the 
 quantum-coherent time evolution of the state of the system described by
unitary transformations which are controlled by the program. 
Elementary steps are 
(i)~the preparation of the initial state of the qubits, 
(ii)~single-bit operations (gates), i.e.\ unitary 
transformation of individual qubit states, triggered by a
modification of the corresponding  
one-qubit Hamiltonian for some period of 
time, (iii)~two-bit gates,
which require controlled inter-qubit couplings, and (iv)~the 
measurement of the final quantum state of the system. 
The phase coherence time has to be
long enough to allow a large number of these coherent processes.
Ideally, in the idle period between the operations the Hamiltonian of
the system is zero to avoid further time evolution of the 
states. 

Several physical realizations of quantum information systems
 have been suggested. Ions in a trap, manipulated by laser
irradiation are the best studied system. However,
alternatives need to be explored, in particular those which are
more easily embedded in an electronic circuit
as well as scaled up to large numbers of qubits. From this point
of view mesoscopic and nano-electronic devices appear particularly
promising\cite{SSH,Loss,Mooij,Averin,Bouchiat}. 
Normal-metal single-electron devices are
discussed in connection with 
classical digital applications and, in fact, constitute 
the ultimate electronic memory~\cite{Likharev-Korotkov}.
However, their use for quantum computation is ruled out, since, 
due to the large number of electron states involved,
different tunneling processes are incoherent.
Ultra-small quantum dots with discrete levels are candidates for
qubits, but their strong coupling to the environment renders their 
phase coherence time short. 
More promising are systems built from Josephson junctions,
where the coherence of the superconducting state can be exploited. 
Quantum extension of elements based on a single-flux logic have
been suggested, and attempts were made to observe
coherent oscillations of flux quanta between degenerate states.
Here we suggest to use low-capacitance Josephson junctions, where 
Cooper pairs tunnel coherently while Coulomb blockade effects allow
the control of the total charge, encouraged by experiments which
demonstrated the superposition of charge states~\cite{Bouchiat,Maassen}.
If biased near degeneracy these junctions constitute qubits 
with two logical states differing by one Cooper-pair charge on an
island.

A simple design of Josephson junctions qubits and their coupling
(reviewed below) has been suggested in  Ref.\ \onlinecite{SSH}. 
Single-bit operations can be performed by controlling  gate
voltages applied to individual junctions, while two-bit gates can be
implemented by tuning the selected qubits to resonance. 
The dephasing time has been estimated to be large
compared to elementary operation
times. To read out the quantum state 
a dissipative normal-metal 
single-electron transistor should be coupled to the  qubit~\cite{SS}. 
The low number of junctions and control voltages of the simple design
should simplify an experimental realization.
Drawbacks are the continuing time evolution of the states 
 also during  idle periods, 
which necessitates a high precision of the time control, as well as
intrinsic errors introduced by nonvanishing two-bit couplings
even if the qubits are out of resonance. 
In this article we suggest an improved design, still
based on nano-scale Josephson junction technology, which is
close to ideal. 
The crucial step is the replacement of 
the Josephson junctions by SQUIDs which can be controlled by
external magnetic fluxes. This allows us to switch the Josephson couplings 
between zero and  non-zero values. It substantially reduces the
requirements on the time control and  provides a
complete control of two-bit couplings.\bigskip

\noindent 
{\bf An ideal model:} To fix ideas we present a model of an ideal
quantum computer with Hamiltonian
\begin{equation}
        H =  - \sum\limits_{i=1}^N \left[ 
                E_z^i(t) \hat\sigma_z^i +  E_x^i(t) \hat\sigma_x^i
                \right]
             + \sum\limits_{i\ne j} A^{ij}(t)  \hat\sigma_+^i \hat\sigma_-^j\; .
\label{idealH}
\end{equation}
A spin notation is used for the qubits 
with Pauli matrices  $\hat\sigma_z$,
$\hat\sigma_x$, $\hat\sigma_{\pm} = (\hat\sigma_x \pm i
\hat\sigma_y)/2$. Ideally,
each energy $E_z^i(t)$, $E_x^i(t)$
and the (real symmetric) couplings $A^{ij}(t)$ can be switched
separately for controlled
times between zero and finite values, $E_z^i$, $E_x^i$ and $A^{ij}$.
We assumed that $E_z^i$ is the largest energy,
suggesting the choice of basis states 
$|\uparrow_i \rangle$ and $|\downarrow_i \rangle$ aligned along the $z$-axis. 
Residual inelastic interactions, which destroy the coherence, 
 and the  measurement device, when turned on,
should be accounted for by
extra terms $H_{\rm res}$ and $H_{\rm meas}(t)$, respectively. \\
(i) For the system  (\ref{idealH}) the  initial state can be prepared
by turning on large values of $E_z^i\gg k_{\rm B}T, i = 1, \dots, N$ at low 
temperature  for sufficient time (while  $E_x^i(t) =  A^{ij}(t) = 0$), 
such that the residual interaction,  $H_{\rm res}$, 
relaxes all spins to the
ground state, $|\uparrow \uparrow \uparrow\dots \rangle$. 
Switching  $E_z^i(t)$ back
to zero leaves the system in a well defined state, and, since
 $H=0$, there is no further time evolution.\\
(ii) Single-bit operations are controlled by turning on one of
the   corresponding  $E_x^i(t)$ for a time $\tau$. 
Hence, the 
spin $i$ evolves according to the unitary transformation
\begin{equation}
\label{1bit_spin_flip}
U_{\rm 1b}^i(\tau)=  
 \exp(-i E_{x}^i \tau \hat\sigma_x^i/ \hbar) \; .
\end{equation}
Depending on the time span, a $\pi/2$- or $\pi/4$-rotation is
performed, producing a spin flip or an equal-weight superposition of
spin states.  Switching on (a small) $E_z^i(t)$ for some time produces another 
needed operation: 
a phase shift between $|\uparrow_i \rangle$ and
$|\downarrow_i \rangle$. \\
(iii) A two-bit operation on qubits $i$ and $j$ is
achieved by turning on the corresponding $A^{ij}(t)$. In the basis  
$|\uparrow_i \uparrow_j \rangle$, 
$|\uparrow_i \downarrow_j \rangle$,
$|\downarrow_i \uparrow_j \rangle$,
$|\downarrow_i \downarrow_j \rangle$ the result is
described by 
\begin{equation}
\label{2bit_Operation}
U_{\rm 2b}^{ij}(\tau)=  
\left( 
\begin{array}{cccc}
1 & 0 & 0 & 0 \\ 
0 & \cos\left({A^{ij} \tau / \hbar}\right)  & 
i\sin\left({A^{ij} \tau / \hbar}\right) & 0 \\ 
0 & i\sin\left({A^{ij} \tau / \hbar}\right) & 
\cos\left({A^{ij} \tau / \hbar}\right)  & 0 \\
0 & 0 & 0 & 1 
\end{array} 
\right) \ .
\end{equation}
For $A^{ij}\tau/\hbar = \pi/2$ the result is a spin-swap operation,
while  $A^{ij}\tau/\hbar = \pi/4$ yields a `square root of swap'. 
The latter transforms the state $|\uparrow_i \downarrow_j \rangle$ into
an entangled state  $\frac{1}{\sqrt{2}} \left(|\uparrow_i \downarrow_j\rangle  
+ i |\downarrow_i \uparrow_j \rangle \right)$.
Combined with the single-bit operations it allows 
to perform a `controlled-not' operation~\cite{Loss}.
This combination provides a universal set of logic gates, sufficient
for quantum computations~\cite{UnivGate,9authors}.\\
(iv) The measurement process has to be discussed for specific
realizations.\bigskip

\noindent 
{\bf  Simple Josephson qubits:}
Nano-scale Josephson junctions can serve as realizations of qubits.
The simplest example is provided by the superconducting electron
box~\cite{SSH} shown in  Fig.\ 1a). The relevant  
conjugate variables are the charge
 $Q=2ne$ on the island (where $n$ is the number of Cooper
 pairs) and the phase difference $\gamma$ across the junction. 
If normal electron tunneling is suppressed by the superconducting
energy gap and only `even-parity' states are involved~\cite{Tinkham},
the circuit dynamics is described by the Hamiltonian
\begin{equation}
        H=\frac{(Q-CV_{\rm x})^2}{2(C+C_{\rm J})} - E_{\rm
        J}\cos\gamma \;\;  ;
        \;\; Q = \frac{\hbar}{i} \,
 \frac{\partial}{\partial(\hbar \gamma/2e)} . 
\label{EnergyQy}
\end{equation}
For the junctions considered,  the charging energy with scale
$E_C \equiv e^2/2(C+C_{\rm J})$ dominates over the Josephson coupling $E_{\rm J}$. 
It is plotted in Fig.\ 2 as a function of the external 
voltage $V_{\rm x}$ for different island charges $n$. In equilibrium at 
$k_{\rm B}T \ll E_C$, the system is in the state
corresponding to the lowest parabola. However, near the voltages
$V_{\rm deg}=(2n+1)e/C$ the states  $n$ and $n+1$
are degenerate, and the Josephson coupling mixes them  strongly. Here,
the reduced two-state Hamiltonian in a basis of the charge states 
$|\uparrow \rangle = |n \rangle$ and  $|\downarrow \rangle = |n+1
\rangle$ is
\begin{equation}
 H= E_{\rm ch}(V_{\rm x}) \hat\sigma_z - \frac{E_{\rm J}}{2}\hat\sigma_x \; ,
\label{MF}
\end{equation}
where $E_{\rm ch}(V_{\rm x}) =\frac{C_{\rm qb}}{C_{\rm J}}e(V_{\rm
x}-V_{\rm deg})$,  and  the capacitance of the qubit in the circuit is $C_{\rm qb}^{-1}=C_{\rm J}^{-1}+C^{-1}$.
\begin{figure}[h]
\epsfysize=10\baselineskip
\centerline{\hbox{\epsffile{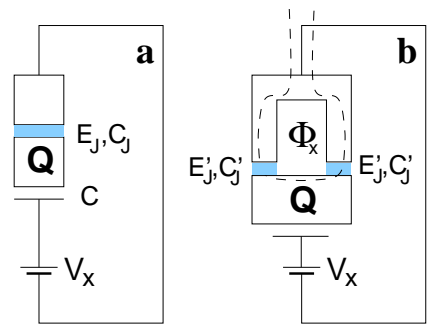}}}
\vskip 0.8cm
\caption[]{\label{Qubit}%
{\sl Josephson junction qubits.}
a) A simple realization of a qubit is provided by the
superconducting electron box. 
The important degree of freedom is the Cooper pair charge $Q=2en$ 
on the island between gate capacitor $C$ and  
Josephson junction (grey area) with capacitance $C_{\rm J}$ 
and Josephson coupling energy $E_{\rm J}$. \\
b) The improved design of the qubit. The island is coupled to the 
circuit via two Josephson junctions with parameters $C'_{\rm J}$ 
and $E_{\rm J}'$. This dc-SQUID can be tuned by the external flux which 
is controlled by the current through the inductor loop (dashed line).
The setup allows switching the effective Josephson coupling to zero.
}
\end{figure}

On the way towards the ideal model
 (\ref{idealH}) we achieved a tunable $E_z^i(t)$, however,
the Josephson coupling is fixed $E_x^i(t) = E_{\rm J}/2$. Still
single-bit operations can be performed by controlling the bias
voltage $V_{\rm x}$~\cite{SSH}. Furthermore, when the qubits
are connected in parallel to a joint inductor (similar as in Fig. 3),
the common $LC$-oscillator mode provides a
two-bit coupling  with  weak, 
but constant $A^{ij} \sim \frac{C^2}{C_{\rm J}^2} \frac{E_{\rm J}^2
 L}{\Phi_0^2}$. 
This coupling provides a two-bit gate
 if two qubits, $i$ and $j$, are brought into resonance by  
biasing them with the same gate voltage $V_{{\rm x}i}= V_{{\rm x}j}$. 
Out of resonance it is only a weak perturbation.

\vskip0.5cm

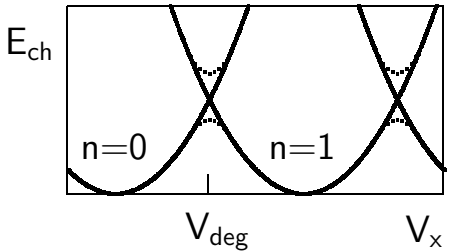
\begin{figure}[h]
\Large
\setlength{\unitlength}{25mm}
\begin{picture}(2,1)(-0.8,-0.15)
\thinlines
\put(-0.25,0){\line(1,0){2}} \put(1.75,0){\line(0,1){1}}
\put(1.75,1){\line(-1,0){2}} \put(-0.25,1){\line(0,-1){1}}
\linethickness{1pt}
\qbezier[1000](-0.25,0.125)(0.228553,-0.353554)(0.707107,1)
\qbezier[1000](0.292893,1)(1,-1)(1.70711,1)
\qbezier[1000](1.29289,1)(1.52145,0.353554)(1.75,0.125)
\put(0,0.25){\makebox(0,0){\sf n=0}} \put(1,0.25){\makebox(0,0){\sf n=1}}
\thinlines
\put(0.5,0){\line(0,1){0.1}}
\put(0.55,-0.2){\makebox(0,0){${\sf V}_{\sf deg}$}}
\put(1.65,-0.2){\makebox(0,0){${\sf V}_{\sf x}$}}
\put(-0.45,0.8){\makebox(0,0){${\sf E}_{\sf ch}$}}
\linethickness{1pt}
\qbezier[10](0.375,0.28125)(0.5,0.5)(0.625,0.28125)
\qbezier[10](1.375,0.28125)(1.5,0.5)(1.625,0.28125)
\qbezier[10](0.375,0.78125)(0.5,0.5)(0.625,0.78125)
\qbezier[10](1.375,0.78125)(1.5,0.5)(1.625,0.78125)
\end{picture}

\vskip 0.8cm
\caption[]{\label{Parabolas}%
{\sl Spectrum of a superconducting electron box.}
The charging energy of the superconducting electron box 
is shown (solid lines) as a function of 
the applied gate voltage $V_{\rm x}$ for different 
numbers of Cooper pair charges $n$ on the island. 
Near degeneracy points the weaker Josephson coupling energy 
mixes the charge states and modifies the energy of the 
eigenstates (dotted line). In this regime the system 
effectively reduces to a 2-state quantum system.
}
\end{figure}

The  external voltage source is part of a dissipative
circuit with  effective resistance $R_V$. This introduces 
fluctuations and destroys the phase coherence.  Following
Refs.\onlinecite{RMP,Weiss}  
we can estimate the corresponding decoherence time.
At the degeneracy point it is 
\begin{equation}
\tau_V= \frac{1}{4\pi}\frac{R_{\rm K}}{R_V} \left(\frac{C_{\rm J}}{C_{\rm qb}}\right)^2 
\frac{\hbar}{E_{\rm J}}
\tanh\left(\frac{E_{\rm J}}{2k_{\rm B}T}\right) \; .
\label{Vdecoh}
\end{equation}
The ratio of the quantum resistance
$R_{\rm K}=h/e^2 \approx 26 \mbox{k}\Omega$ and $R_V$   determines the
strength of the fluctuations.  Furthermore,
a small gate capacitance $C \ll C_{\rm J}$ helps decoupling the
qubit from the environment.
Both should be optimized to provide a phase coherence time
much longer than typical operation times $\hbar/E_{\rm J}$.

A problem with the simple design is that the
eigenstates of the Hamiltonian  
(\ref{MF}) are non-degenerate at all voltages. Therefore, the 
relative phase of two eigenstates evolves in time even during idle
periods. 
We can still store  quantum information in the qubit, as
becomes apparent after a transformation to the interaction
representation. This introduces, however, an explicit
time dependence in the operators with nontrivial consequences
for the unitary transformations. Their result does not only depend  on 
the time span $\tau$ of the operations  but also on the time $t_0$ when they 
start. As a consequence the time
elapsed since the beginning of the computation should be
controlled with high accuracy, determined  
by the spacing between the two eigenvalues of $H(V_0)$.
A second problem of the simple design is 
the non-vanishing two-bit coupling even out of resonance.
It introduces an error in the computation.  The design  discussed
below overcomes both these problems.\bigskip

\noindent 
{\bf  Josephson qubit with SQUID-controlled coupling:}
The crucial step towards the ideal model 
(\ref{idealH}) is to make the Josephson coupling tunable. This is
achieved by the design shown in Fig.\ 1b), where 
each Josephson junction is replaced by a dc-SQUID threaded by the flux $\Phi$.
The SQUID  is biased by an external flux $\Phi_{\rm x}$, which is  coupled
into the system through an inductor loop. The energy of this element is
\begin{eqnarray}
\nonumber
        E_{\rm SQUID}= & \frac{1}{2L_\Phi} (\Phi-\Phi_{\rm x})^2 - 2E'_{\rm J}\cos\gamma \,
        \cos\left(\frac{\pi\Phi}{\Phi_0}\right) \\
         & + C'_{\rm J} (\frac{\hbar^2}{4e^2}\dot\gamma^2+\frac{1}{4}\dot\Phi^2) \; ,
\end{eqnarray}
where $\Phi_0=h/(2e)$.
The phase difference across the element $\gamma$ and 
flux $\Phi$ are dynamical variables. 
If  the self-inductance $L_\Phi$ of the loop is low~\cite{Tinkham},
 $\Phi_0^2/L_\Phi \gg (2\pi)^2E'_{\rm J}, e^2/C'_{\rm J}$, 
fluctuations of the flux around $\Phi_{\rm x}$ are weak. 
Furthermore, if the frequency of flux oscillations
$\omega_\Phi =(L_\Phi C'_{\rm J}/2)^{-1/2}$ is large, 
$ \hbar \omega_\Phi \gg E'_{\rm J}, E_{\rm ch}, k_{\rm B}T$, the $\Phi$-degree of
freedom is in the ground state. In this case the SQUID-controlled
qubit is described by a Hamiltonian of the form
(\ref{EnergyQy}) with potential energy 
\begin{equation}
        - 2E'_{\rm J}\cos(\pi\Phi_{\rm x}/\Phi_0)\cos\gamma
\label{EJPhi}
\end{equation} 
and effective junction  capacitance $C_{\rm J}=2C'_{\rm J}$.
I.e., the effective Josephson coupling
$E_{\rm J}(\Phi_{\rm x})=2E'_{\rm J}\cos(\pi\Phi_{\rm x}/\Phi_0)$ is  tunable by the
external flux $\Phi_{\rm x}$ between
$2E'_{\rm J}$ and zero. 

We note that the SQUID-controlled qubit is described by the first two
terms of the ideal Hamiltonian (\ref{idealH}), with  $z$- and $x$-components
 controlled independently by the gate voltage and the 
 the flux. In the idle state we keep $V_{\rm x}=V_{\rm deg}$ and 
$\Phi_{\rm x}=\Phi_0/2$ so that the Hamiltonian $H=0$. If we change one of
them the new Hamiltonian generates rotations around $z$- 
or $x$-axis, respectively, which are elementary one-qubit operations.
Note, that with this design there is no need to control the total
operation time $t_0$, while the voltage and $E_{\rm J}(\Phi_{\rm x})$ 
can be optimized such that the duration of the manipulations $\tau$ is
long enough to simplify 
time control and short enough to speed up the computation.

The circuit of the current source with resistance  $R_I$,
which couples the flux $\Phi_{\rm x}$ to the SQUID
by the  mutual inductance $M$, introduces
fluctuations and may destroy the 
coherence of the qubit dynamics. At the degeneracy point the decoherence
time is
\begin{equation}
\tau_I= \frac{1}{\pi^3} \frac{R_I}{R_{\rm K}}
\left(\frac{\Phi_0^2}{E'_{\rm J}M}\right)^2 \frac{\hbar}{k_{\rm B}T} \; .
\label{Idecoh}
\end{equation}
This dephasing is slow if the current source is
 coupled weakly to the qubit (small $M$) and its  resistance  is high.\bigskip

\noindent 
{\bf  Controlled inter-qubit coupling:}
The controlled Josephson junctions allow us also to switch the
two-qubit interaction for each pair of qubits, bringing us close to the
ideal model (\ref{idealH}).
The simplest implementation of the coupling is to connect
all $N$ qubits in parallel to each other to an inductor $L$ (Fig.\ 3).
If the frequency  of the $LC$-mode in the resulting circuit,
$\omega_{LC}=(NC_{\rm qb}L)^{-1/2}$, is large $\hbar\omega_{LC} \gg
E_{\rm J}, E_{\rm ch}, k_{\rm B}T$  the fast
oscillations produce an effective coupling of the qubit dynamics \cite{SSH} 
\begin{equation}
        H_{\rm int}
        =\sum\limits_{i<j}\frac{E_{\rm J}^iE_{\rm J}^j}{E_L}\hat\sigma_y^i\hat\sigma_y^j ,
\label{HintGen}
\end{equation}
where
$E_L=\frac{\Phi_0^2}{\pi^2L} \left(\frac{C_{\rm J}}{C_{\rm qb}}\right)^2$,
and $E_{\rm J}^i= E_{\rm J}(\Phi_{{\rm x}i})$ are the effective Josephson energies of the 
qubits, controlled by the external fluxes.
The coupling energy (\ref{HintGen}) can easily be understood as 
the magnetic energy of the current in  
the inductor, where the current is the sum of contributions from all
qubits
$I^i \propto E_{\rm J}^i \hat \sigma_y^i$.

\begin{figure}[h]
\epsfysize=10\baselineskip
\centerline{\hbox{\epsffile{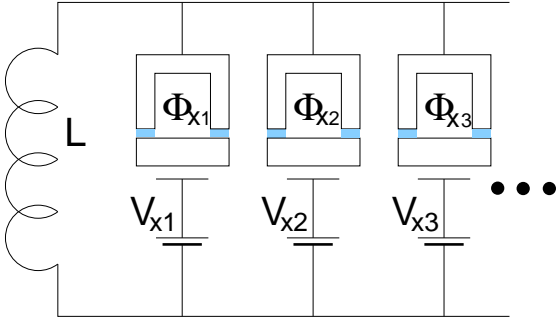}}}
\vskip 0.8cm
\caption[]{\label{Computer}%
{\sl Design of a quantum computer.}
The coupling of the qubits is provided by the $LC$-oscillator mode in 
circuit shown. Note that the system can be scaled to large numbers 
of qubits. In the idle state all effective Josephson couplings are tuned 
to zero and the voltages are chosen such that the charge states are 
degenerate. Single-bit operations are achieved by changing the gate
voltage or flux of one qubit at a time. 
Two-bit operations between any two qubits are triggered by turning 
on the corresponding two Josephson couplings.
}
\end{figure}

Using this interaction we can perform all gate operations. In the
idle state the interaction Hamiltonian  
(\ref{HintGen}) is zero since all the Josephson couplings are turned
off (the charge $Q_i$ is conserved and the 
qubit does not contribute to the current through the inductor). The 
same is true during a one-qubit operation, as long as we perform   
one such an operation at a time: i.e.\ only one $E_{\rm J}^i \ne 0$.
To perform a two-qubit operation with any given pair of qubits, say 1 and 2,
$E_{\rm J}^1$ and $E_{\rm J}^2$ are switched on simultaneously, yielding
the total Hamiltonian 
\begin{equation}
H =- \frac{E_{\rm J}^1}{2}\hat\sigma_x^1 - \frac{E_{\rm J}^2}{2}\hat\sigma_x^2 + 
\frac{E_{\rm J}^1E_{\rm J}^2}{E_L}\hat\sigma_y^1\hat\sigma_y^2 \; .
\end{equation}
While not identical to the form (\ref{idealH}) also these  two-bit
 operations, combined with the one-bit operations discussed above, 
provide a complete set of gates  required  for 
quantum computation~\cite{UnivGate}.\bigskip

\noindent 
{\bf  Discussion and outlook:} 
To demonstrate that the constraints on the circuit
parameters can be met by available technology, we suggest a
suitable set: \\
(i) We choose  junctions with capacitance $C_{\rm J}=3\cdot 10^{-16}$F, corresponding to
a charging energy  (in temperature units) $E_C \sim 3$K, and  a 
smaller gate capacitance $C=3\cdot 10^{-17}$F to reduce the coupling to the
environment. The superconducting gap has to be slightly larger $\Delta
> E_C$. Thus at working temperature of order
$T=50$mK the initial thermalization is assured. 
We further choose $E'_{\rm J}=50$mK, i.e.\ the time scale of
one-qubit operations is $\tau_{\rm op}=\hbar/E_{\rm J}\sim 
7\cdot 10^{-11}$s. Fluctuations associated with the gate
voltages (\ref{Vdecoh}), with resistance $R_V\sim 50\Omega$,
 limit the coherence time to
$\tau_V/\tau_{\rm op}\sim 4000$ operations.\\
(ii) If the inductance of the SQUID loop is $L_\Phi=0.1$nH, 
fluctuations of the flux  
are weak, $\langle \delta\Phi^2\rangle^{1/2} \sim 0.08 \Phi_0$, and,
as required, the excitation energy
of the $\Phi$-degree of freedom $\hbar \omega_\Phi\sim 
80$K is large compared to characteristic energy scales of the qubit.
Compared to the voltage fluctuations, 
 for reasonable values $M=1$nH and $R_I=10^2$--$ 10^6\Omega$, the flux
circuit has a weak dephasing effect.\\
(iii) To assure fast  two-bit operations we choose the energy scale $E_L$
of the order of $10E_{\rm J}$, which is achieved for 
 $L\sim 3\mu$H. Since the energy of $LC$-oscillations
$\hbar(NC_{\rm qb}L)^{-1/2}$  
should be large compared to $k_{\rm B}T$ and $2E'_{\rm J}$, the number of qubits
in the circuit is limited by $N_{\rm max}\sim 70$.

We add two technical remarks:\\
(i) Above we assumed a very small SQUID inductance $L$. For finite values
 fluctuations of the flux renormalize the energy (\ref{EJPhi}). But
still, by symmetry arguments, at $\Phi_{\rm x}=\Phi_0/2$  
the effective Josephson coupling is zero.\\
(ii) While the expression (\ref{HintGen}) is valid in leading order in an 
expansion in $E_{\rm J}^i/\hbar\omega_{LC}$,  higher terms
also vanish  when the Josephson couplings are put to zero. Hence, the
decoupling in the idle periods persists.

Further remarks on the design and manipulation of the system  are in order:\\
(i) The two lowest states of the qubit are separated from higher states,
which exist in the physical system, by the 
 energies $E_C, \hbar\omega_{LC}, \hbar \omega_\Phi$. 
 If, in addition,
switching processes of $V_{\rm x}$ and  $\Phi_{\rm x}$ are slow on 
the corresponding time scales, the requirements presented above also ensure that 
the higher states are not excited.
Alternatively, instead of sudden switching, one can apply resonance ac-signals
after changing the biases adiabatically to finite values.\\
(ii) In addition to the gate operations the 
resulting quantum state has to be read out. 
This can be accomplished by
coupling a normal-state single-electron transistor capacitively to a
qubit. The important aspect is that
during the computation the transistor is kept in a zero current state
and adds only to the total capacitance. When the transport voltage is
turned on, the
dissipative current in the transistor depends on the state of
the qubit, and the phase coherence of the q-bit is destroyed. This
quantum measurement process has been described explicitly in 
Ref. \onlinecite{SS} by an analysis of  the time-evolution of the density matrix
of the coupled system.\\
(iii) The system presented here does not permit parallel operations on different 
qubits, which is an essential element of many powerful quantum
algorithms. It  can be 
achieved in principle by a more advanced design, making use, e.g.,
of further  tunable SQUIDs decoupling different parts of the
circuit. Such modifications, as  
well as further progress of nano-technology, will provide longer
coherence times  and allow scaling to larger numbers
of qubits. We stress, however, that many aspects
of quantum informations processing can initially be tested on 
simple circuits as proposed here.

To conclude, the realization of a nano-scale quantum computers based on
controlled   Josephson qubits is possible with current technology. In such a 
system fundamental features of macroscopic quantum-mechanical systems can be 
further explored.

{\bf Acknowledgments:}
We thank T.~Beth, M.~Devoret, D.~P.~DiVincenzo, E.~Knill, K.~K.~Likharev, and 
J.~E.~Mooij for stimulating discussions.\medskip

Correspondence should be addressed to Y.M. (e-mail: makhlin@tfp.physik.uni-karlsruhe.de)


\noindent
\begin{thebibliography}{99}


\bibitem{SSH}
Shnirman,~A., Sch\"on,~G., and Hermon,~Z. Quantum manipulations of small 
Josephson junctions. {\it Phys. Rev. Lett.} {\bf 79}, 2371 (1997).

\bibitem{LloydIntro} Lloyd,~S. A potentially realizable quantum computer.
{\it Science} {\bf 261}, 1589 (1993).

\bibitem{Bennett} Bennett,~C.~H. Quantum information and computation.
{\it Physics Today} {\bf 48} (10), 24 (1995).

\bibitem{DV} DiVincenzo,~D.~P. Quantum computation. {\it Science} {\bf 269}, 255 
(1995).

\bibitem{Barenco} Barenco,~A. Quantum physics and computers. {\it Contemp. 
Phys.} {\bf 37}, 375 (1996).

\bibitem{Loss} Loss,~D., DiVincenzo,~D.~P. Quantum computation with quantum 
dots. {\it Phys. Rev. A} {\bf 57}, 120 (1998). 

\bibitem{Mooij} Mooij,~J.~E. private communication.

\bibitem{Averin} Averin,~D.~V. Adiabatic quantum computation with Cooper pairs. 
{\it Solid State Commun.} {\bf 105}, 659 (1998).

\bibitem{Bouchiat}
Bouchiat,~V., Joyez,~P., Esteve,~D., and Devoret,~M.
to be published in Physica Scripta; V.~Bouchiat, 
Ph.~D. Thesis, Universit\'e Paris 6, (1997).

\bibitem{Likharev-Korotkov}
Korotkov,~A.~N., Chen,~R.~H., and Likharev,~K.~K. Possible performance of 
capacitively coupled single-electron transistors in digital circuits.
{\it J.\ Appl.\ Phys.}\ {\bf 78}, 2520 (1995).

\bibitem{Maassen}
Maassen~v.d.~Brink,~A., Sch\"on,~G., and Geerligs,~L.~J. Combined 
single-electron and coherent-Cooper-pair tunneling in voltage-biased Josephson 
junctions. {\it Phys.\ Rev.\ Lett.}\ {\bf 67}, 3030 (1991).


\bibitem{SS}
Shnirman,~A. and Sch\"on,~G. Quantum measurements performed with a 
single-electron transistor. {\it Phys. Rev. B} {\bf 57}, 15400 (1998).

\bibitem{UnivGate} Lloyd,~S. Almost Any Quantum Logic Gate is Universal. {\it 
Phys. Rev. Lett.} {\bf 75}, 346 (1995).

\bibitem{9authors} Barenco,~A. et al. Elementary gates for quantum computation. 
{\it Phys.\ Rev.\ A} {\bf 52}, 3457 (1995).

\bibitem{Tinkham} see, e.g., Tinkham,~M. Introduction to Superconductivity. 
McGraw-Hill, New York, 1996.

\bibitem{RMP} Leggett,~A.~J. et al. Dynamics of the dissipative two-state 
system. {\it Rev. Mod. Phys.} {\bf 59}, 1 (1987).

\bibitem{Weiss} Weiss,~U. Quantum dissipative systems. World Scientific, 
Singapore, 1993.

\end{thebibliography}
\end{document}